\documentclass[times, 10pt,twocolumn]{article} 
\usepackage{latex8}
\usepackage{times}
\usepackage{graphicx}
\usepackage{multirow}
\usepackage{array}
\usepackage{tikz}
\usepackage{adjustbox}
\usepackage{makecell}

\usepackage{seqsplit}
\newcommand{\urll}[1]{{\seqsplit{#1}}}

\newcolumntype{P}[1]{>{\centering\arraybackslash}p{#1}}
\newcolumntype{C}[1]{>{\centering\let\newline\\\arraybackslash\hspace{0pt}}m{#1}}

\begin{document}
\title{DNS Covert Channel Detection via Behavioral Analysis: a Machine Learning Approach}

\author{Salvatore Saeli, Federica Bisio, Pierangelo Lombardo, Danilo Massa\\
aizoOn Technology Consulting\\ Strada del Lionetto 6, 10146 Turin, Italy\\ \{salvatore.saeli,federica.bisio,pierangelo.lombardo,danilo.massa\}@aizoongroup.com\\
}

\maketitle
%
%-------------------------------------------------------------------------------
\begin{abstract}
%-------------------------------------------------------------------------------

Detecting covert channels among legitimate traffic represents a severe challenge due to the high heterogeneity of networks.
Therefore, we propose an effective covert channel detection method, based on the analysis of DNS network data passively extracted from a network monitoring system. 
The framework is based on a machine learning module
and on the extraction of specific anomaly indicators able to describe the problem at hand.
The contribution of this paper is two-fold: (\emph{i}) the machine learning models encompass network profiles tailored to the network users, and not to the single query events, hence allowing for the creation of behavioral profiles and spotting possible deviations from the normal baseline; (\emph{ii}) models are created in an unsupervised mode, thus allowing for the identification of zero-days attacks and avoiding the requirement of signatures or heuristics for new variants.
The proposed solution has been evaluated over a 15-day-long experimental session with the injection of traffic that covers the most relevant exfiltration and tunneling attacks: all the malicious variants were detected, while producing a low false-positive rate during the same period.
\end{abstract}

%-------------------------------------------------------------------------------
\section{Introduction}
%-------------------------------------------------------------------------------
% 
One of the most serious threats to the current society is represented by cybercrime, with heavy and sometimes dramatic consequences for many companies, organizations and single individuals~\cite{grady2005cyber,grance2004cybersec,korolov2012cyber,taylor2014cyber,yadav2016cyber_sec}. Data exfiltration, in particular, plays a key role in the cybercrime scenario, as it is related to the stealing of sensitive information~\cite{giani2006data}.
Among different techniques of exfiltration, \emph{covert channels} represent a significant threat for defenders, as they are widely used and their detection is challenging.
A covert channel~\cite{zander2007survey} can be defined as a way to communicate, transfer or exfiltrate data while exploiting network resources never intended for this purpose. The aim of such a technique is to extract sensitive information from organizations and companies, while eluding conventional security measures (e.g., intrusion detection systems and firewalls).

Among the reasons that make covert channels a severe menace for threat hunters, it is worth mentioning the following: (\emph{i}) conventional intrusion detection and firewall systems typically fail in the detection of covert channels; (\emph{ii}) the network traffic varies considerably, thus causing detection issues for classical statistical approaches to covert channel detection; (\emph{iii}) related to the previous two points, another issue is the difficulty in distinguishing covert channels between legitimate communications, this is often caused by the absence of focus on users behavioral analysis; (\emph{iv}) although many works focus on the process of tunnel attacks, only a very restricted number of them analyzes the properties useful to describe the data exfiltration process. 

In this paper, we propose an effective technique for the detection of DNS covert channels, based on the analysis of network data passively extracted by a network monitoring system. The proposed framework is based on a machine learning module and on the extraction of specific anomaly indicators able to describe the problem at hand. 
The focus of the machine learning module is the creation of models embedding the behavioral characteristics of a user and hence spotting possible deviations from the normal baseline.
The power of this approach is the ability to identify zero-days attacks, without the requirement of signatures or heuristics for new variants. However, it may carry the drawback of highlighting also non-malicious events that are similar to covert channels from the perspective of DNS behavior. This could lead to a large number of false positives, but it is mitigated by a subsequent \emph{advanced analytics} module, which encodes cyber security knowledge. This structure allows the proposed framework to identify a dual typology of attacks: (\emph{i}) \emph{exfiltration}, and (\emph{ii}) \emph{tunneling}.

The remainder of the paper is structured as follows. Sect.~\ref{sec:work} provides an overview of the related work. %Sect.~\ref{sec:ml} describes the possible advantages provided by machine learning to the covert channel detection problem.
In Sect.~\ref{sec:aramis} we briefly describe the monitoring platform containing the covert channel detection method, which is the focus of this paper. After an introduction on the problem, given in Sect.~\ref{sec:pf}, the proposed approach is described thoroughly in Sect.~\ref{sec:cc}. Finally, we discuss the experimental results in details in Sect.~\ref{sec:res} and possible future developments in Sect.~\ref{sec:concl}.

%-------------------------------------------------------------------------------
\section{Related Work}\label{sec:work}
%-------------------------------------------------------------------------------
State-of-the-art works regarding covert channels usually distinguish between two different types: (\emph{i}) storage covert channels~\cite{cabuk2009ip}, where covert bits are strictly bounded to the communication protocols under analysis (e.g., IP, DNS, HTTP, SMB, SSL); (\emph{ii}) timing covert channels~\cite{shrestha2016support}, based on the manipulation of timing or on the ordering of network events (e.g., packet arrivals).

Depending on the type of covert channel, associated detection technique may also vary: (\emph{i}) for the storage covert channels, typical detection methods involve Markov Chains and Descriptive Analytics~\cite{butler2011quantitatively}; (\emph{ii}) for the timing covert channels, many different approaches have been considered, in particular: statistical tests of traffic distribution~\cite{paxson2013practical}, regularity tests of time variations within the traffic~\cite{ellens2013flow}, and machine learning methods~\cite{satam2015anomaly} such as Support Vector Machines (SVMs)~\cite{aiello2015dns} and Bayesian Networks~\cite{aiello2014supervised}.

The detection framework proposed in this article focuses on storage covert channel, while borrowing some detection techniques typically used for the detection of timing covert channels, e.g., the use of SVMs. For the proposed technique, DNS covert channels are taken into consideration. This is motivated by the fact that, at the time of this writing, DNS represents one of the most common protocol to control systems and exfiltrate data covertly~\cite{blackhat2015,rootedcon2012,rsaconf2018}. Nevertheless, DNS covert channel is a method that many organizations still fail to detect. In its simplest form, this technique employs the DNS protocol to communicate directly with an attacker's external DNS server.

DNS is not designed to exchange an arbitrary amount of data: therefore, messages are usually short and answers are not correlated and may not be received in the same order as the corresponding requests. Usually attackers avoid these limitations with two approaches: (\emph{i}) \emph{tunneling}~\cite{liu2017detecting}: the attacker establishes a bidirectional channel to send communications and instructions from an external server (command and control, or C\&C) to a compromised host; (\emph{ii}) \emph{exfiltration}~\cite{nadler2019detection}: the attacker executes data exfiltration from a compromised host towards a controlled external server, sending information with minimal overhead and short and independent requests.

Ahmed et al.~\cite{ahmed2018real} analyzed data extracted by a network monitoring system, as several other works in state-of-the-art~\cite{aiello2015dns,kara2014detection,marchal2012dnssm}. The authors used only stateless attributes of individual DNS queries, based on three main categories as characters count (e.g., total count of characters in FQDN, count of characters in subdomains, count of uppercase and numerical characters), entropy on strings, and length of discrete labels in the query (e.g., maximum label length and average label length). The authors developed an anomaly detection system based on Isolating Forest (iForest). %Nevertheless, t
The approach does not involve any particular DNS record type, the test phase is deployed solely with the exfiltration tool DET and the training phase of the model is computationally very expensive.

Nadler et al.~\cite{nadler2019detection} proposed an approach based on the characteristics of the queries employed for data exchange over DNS (e.g., longer than the average requests and responses, with encoded payload and a plethora of unique requests) and on the use of single domains for exfiltration. As in the previous case, the anomaly detection was based on iForest. However, this approach considers only A and AAAA records usage, the tested malware was simulated with ad-hoc crafted queries in experimental environment; iodine and dns2tcp were the only tunneling tools taken into account.

Das et al.~\cite{das2017detection} underlined that DNS tunneling and malware exfiltration typically involve an exchange with the attacker server of a part of payload encoded in a subdomain portion of the DNS query or in the response packet. The authors proposed two machine learning approaches, (\emph{i}) a logistic regression model and (\emph{ii}) a k-means clustering, respectively (\emph{i}) for the exfiltration and (\emph{ii}) for the tunneling scenarios; these approaches are based on grammatical features extracted from queries representative of an encoded payload (e.g., entropy and number of upper cases, lower cases, digits, and dashes characters). However, the test phase lacks several attack scenarios; in particular, for DNS tunneling, only the tunneling tool dnscat2 with the TXT record was employed.

Liu et al.~\cite{liu2017detecting} analyzed the traffic of several open-source DNS-tunneling tools based on the extraction of four kinds of features, including time-interval features (e.g., mean and variance of time-intervals between a request and a response), request packet size, domain entropy, and distinction of record types (e.g., A, TXT, MX). The authors developed a binary supervised classification model based on the description of traffic generated from different DNS-tunneling tools (e.g., dnscat2, iodine, and dns2tcp). The solution proposed by \cite{liu2017detecting} consists of an offline component, where the system trains the classifier, and an online component, where the system identifies the tunnel traffic. Nevertheless, during the training phase the DNS traffic is not collected in a streaming fashion but using a specific dataset.
%

%Network Monitoring Tool
\section{Network Monitoring Platform}\label{sec:aramis}

The proposed covert channel detection algorithm has been deployed in a network security monitoring platform called \emph{aramis} (Aizoon Research for Advanced Malware Identification System) able to automatically identify a wide range of malware and attacks in near-real-time. This software is bundled with dedicated hardware\footnote{E5-2690 2.9GHz x 2 (2 sockets x 16 cores) 16 x 8GB RAM, 1.1TB HDD}, and its structure can be summarized in four phases:

\begin{enumerate}
\setlength\itemsep{0em}
\item Collection: sensors are placed in various nodes of the network. Each sensor gathers the data from its segment of the network, pre-analyzes them in real-time and sends the results to a NoSQL database.
\item Enrichment: inside the NoSQL database, data is enriched with information coming from the Cloud Service, which collects intelligence from various OSINT sources and from internally managed sources.
\item Analysis: two kinds of analyses are executed on the stored data: (i) \emph{advanced
cybersec analytics} to spot and highlight specific attacks, among which the covert channel detection module can be found, and (ii) a \emph{machine learning engine} which compares the behavior of each node with the usual one.
\item Visualization: the results are presented through cognitive dashboards, crucial to highlight anomalies.
\end{enumerate}

%-------------------------------------------------------------------------------
\section{A Glimpse into DNS Covert Channels}\label{sec:pf}
%-------------------------------------------------------------------------------
In DNS covert channel technique the main challenge is represented by performing the data exchange in an optimized manner. To accomplish this goal, the structural and grammatical characteristics of hostname and the capabilities of various DNS record types are exploited~\cite{sans2013}. According to the RFC 1034, the hostname can be up to 253 characters long and it is composed by labels each of which can be up to 63 characters long; each character can be a letter (upper case and lower case are both permitted), a number or a hyphen. 
To maximize the data exchanged - the payload transmitted - in each DNS query, each label of the subdomain contains a portion of the data, which have previously been encoded (e.g., using Base32, Base64, Base128 or Hexadecimal codecs) or encrypted (e.g., using RC4 encryption algorithm). Furthermore, depending on the DNS covert channel scenario, specific DNS record types are employed. We distinguish between two different cases: (\emph{i}) \emph{exfiltration}: information might be contained in a subdomain of a domain and usually the communication is based on A or AAAA DNS record types; (\emph{ii}) \emph{tunneling}: a bidirectional channel between a compromised host and a server controlled by the attacker is established in order to send commands and obtain information about the host. The requirement to maintain an open connection determines the choice of  certain DNS records -- such as TXT (which is the most common choice), KEY, CNAME, MX, SRV, NULL, and PRIVATE -- that allow transmitting arbitrary portions of text, in addition to the information exchanged over the subdomain.

Another crucial point is represented by the information stored in the stub revolver cache. In fact, normal DNS traffic is usually reduced by the fact that a huge amount of DNS responses are cached within the stub resolver; instead, when a covert channel takes place, the domain-specific traffic tends to avoid cache by using non-repeating or short time-to-live messages, resulting in not repeated and unique queries~\cite{sans2013}.

%-------------------------------------------------------------------------------
\section{Detection of DNS Covert Channels}\label{sec:cc}
%-------------------------------------------------------------------------------

The proposed DNS covert channel detection technique is based on the monitoring of large volumes of DNS requests from a given IP address and on the analysis of the domain-related linguistic features~\cite{bisio2017real}. The main idea is to employ machine learning to create models able to embed the behavioral characteristics of network users' traffic. %An event is considered abnormal when it results distant from the typical behavior modeled by the algorithm.
An event is considered anomalous when its results are distant from the typical behavior modeled by the algorithm. Note that the analysis relates different queries, and therefore the models encompass network profiles tailored to the network users, and not to the single query events. Furthermore, the possible risk of false positives generation related to the use of machine learning %(see Sect.~\ref{sec:ml}) 
is mitigated by the analysis of specific descriptors of the DNS tunneling or exfiltration process built out of the network data.

\begin{figure}[!ht]
\centering
\hspace*{-0.5cm}
\includegraphics[width=9.2cm]{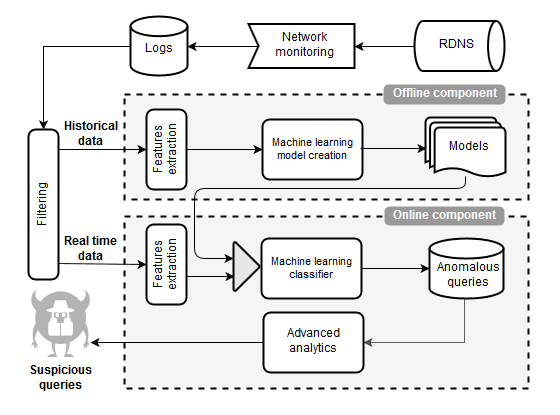}
\caption{\label{fig:workflow} Diagram  representing the main steps of the detection algorithm: (\emph{\textbf{i}}) filters, (\emph{\textbf{ii}}) offline phase, and (\emph{\textbf{iii}}) online phase. 
}
\end{figure}

The general process of the resulting framework is shown in Figure~\ref{fig:workflow}. The raw local recursive DNS server (RDNS) data are collected, parsed and transformed into logs by the network monitoring platform described in Sect.~\ref{sec:aramis}. Logs are filtered as described in Sect.~\ref{sec:filters}, both in the offline and in the online component: the first one periodically%
\footnote{The period was set to six hours as a tradeoff between the need to analyze enough data to build reliable models
and the need to avoid the abuse of hardware resources.}
extracts the features described in Sect.~\ref{sec:offline} from the historical data collected during the period between two subsequent runs and builds the related machine learning models. These models are then used by the online component, which extracts the related features, detects anomalous queries in real time, and further analyzes them in order to spot suspicious queries, which will be the final outputs of the algorithm, as described in Sect.~\ref{sec:online}.
In the following, the three main phases that compose the algorithm are summarized: \emph{(i) filters}: filtering of input RDNS queries extracted by a network analyzer, \emph{(ii) offline phase}: assessment of the network under analysis and creation of models able to describe the normal behavior of the network, \emph{(iii) online phase}: validation of such models to spot potentially anomalous and/or malicious activities occurring in the network.

%---------------------------
%% %---------------------------

%-------------------------------------------------------------------------------
\subsection{Filters}\label{sec:filters}
%-------------------------------------------------------------------------------
The input for both phases is represented by DNS queries passively extracted by the network analyzer described in Sect.~\ref{sec:aramis}.
In order to consider the
traffic that may actually be associated with exfiltration or tunneling, the following DNS query types are analyzed: TXT, CNAME, MX, SRV, NULL, KEY, A, AAAA. 
The collected data are filtered as described in Table~\ref{tab:filters}: any query or group of queries matching any of the filters is removed from the subsequent analysis.

%-------------------------------------------------------------------------------
\subsection{Offline Phase}\label{sec:offline}
%-------------------------------------------------------------------------------
The offline phase takes as input the historical data collected from the network and representing a baseline of the normal behavior. This phase includes:

\begin{enumerate}
\setlength\itemsep{0em}
\item \emph{Feature extraction}: raw features are parsed in order to extract the valuable information useful to describe the problem at hand.
\item \emph{Model  creation}: a classifier is trained using the extracted features, with the aim of distinguishing between patterns performing with a normal behavior and potentially suspicious patterns.
\end{enumerate}

%
%
%%%%%
\renewcommand{\arraystretch}{1.1}
\begin{table}[!ht]
\large
\centering
\caption{Filters Description}
\label{tab:filters}
\begin{adjustbox}{max width=0.48\textwidth}
\begin{tabular}{ |C{3.2cm} | C{9.0cm}|}
\hline
\textbf{Type}                                                                           & \textbf{Description} \\ \hline
White list domains                                                                      & Domains known to be trusted, e.g., the top 10000 domains in the world provided by Alexa~\cite{alexa} and the web URLs of the 500 world biggest companies provided by Forbes~\cite{forbes}. \\ \hline
Response code                                                                           & Only communications which occurred with no error (associated with response code=0) are kept and the analysis of retransmissions due to network errors is avoided.  \\ \hline
Content Delivery Networks                                                               & These represent an important source of false positives for DNS-based detection algorithms~\cite{lombardo2018fast}.    \\ \hline
Overloaded DNS                                                                          & DNS queries are often overloaded so to provide anti-spam or anti-malware techniques.       \\ \hline
Local and corporate domains	                                                            & These domains represent a high percentage of the legitimate  DNS traffic in a corporate network.   \\ \hline
IP addresses in the
subdomain                                                                               & Domains containing an IP address in the 
subdomains 
are typically associated with answers to PTR requests.
\\ \hline
Longest label with less than 6 characters                                                   & We assume that the longest labels used to transfer a payload necessitate of a minimum number of characters.
\\ \hline
Less than 3 different hostnames per domain                                           & After grouping queries with the same domain, we select domains associated with at least 3 distinct hostnames, in order to discard the ones which cannot be involved in an exfiltration process (due to the minimum number of queries involved in the process).
\\ \hline
Duplicated queries                                                                      & Duplicated queries 
are mainly related with retransmissions, quite common in complex networks.  \\ \hline
\end{tabular}
\end{adjustbox}
\end{table}
\renewcommand{\arraystretch}{1}
%%%%%
%
%

\subsubsection{Feature Extraction}

The features used for training the classifier are the following.

\emph{Uppercase Characters Ratio.}
Ratio between the number of uppercase characters and the total number of characters in the 
subdomain. A large fraction of uppercase characters may be associated with a payload encoded in the subdomain using Base64 encoding,
which is widely employed in 
DNS covert channels.

\emph{Digits Ratio.}
Ratio between the number of digits and the total number of characters in the subdomain. A large fraction of digits may indicate a payload encoded or encrypted into the subdomain.

\emph{Total Label Ratio.}
Total number of characters of the hostname divided by 253 (maximum number of characters allowed for a hostname according to RFC 1034). In a wide variety of DNS covert channel scenarios, a payload is encoded or encrypted in the subdomain, aiming at transferring the maximum possible amount of data to conduct the attack. As a result, the obtained queries possess a larger number of characters compared with legitimate queries.

\emph{Per Label Ratio.}
Number of characters of the longest label of the subdomain divided by 63 (maximum number of characters allowed for each label in the hostname according to RFC 1034). The encoded or encrypted payload might contain several types of information (e.g., bot-id, campaign-id, command), in addition to the data exchanged. As a result, one or more labels, used to exchange data, may
contain a large fraction of the characters in the subdomain, while in ordinary traffic the lengths of the labels are typically comparable, resulting in a moderate length for the longest label.

\subsubsection{Model Creation}

A one-class Support Vector Machine (SVM)~\cite{scholkopf2000support} classifier with radial basis function kernel has been used to create the models out of the feature space described in the previous subsection. In fact, even though the original formulation of SVMs is related to the resolution of supervised tasks, the one-class SVM -- which has been shown to be an appropriate choice in the context of anomaly detection \cite{swersky2016evaluation} -- is defined as a boundary-based anomaly detection method, which modifies the original SVM approach by extending it in order to deal with unsupervised data. In our context, this means that the proposed approach is able to train the classifier by using only
the normal network traffic, preserving the malicious samples for the test of the algorithm. In particular, this implies that the proposed method is more likely to identify a new variant of a DNS covert channel, as it does not require a specific training for that variant. %In order to do so, one class SVM encloses the class present in the training set in the optimal volume hyper-sphere~\cite{swersky2016evaluation}.

Like traditional SVMs, one-class SVMs can also be extended to non-linearly transformed spaces using the so called kernel trick, which amounts to define an appropriate scalar product in the feature space.
In the present work a \emph{radial basis function} kernel has been used for the reasons described in \cite{keerthi2003asymptotic}, i.e., the  scalar product between two features vectors $\vec{x}$ and $\vec{x'}$ has been defined as in Eq.~\ref{eq:kernel}
\begin{equation}\label{eq:kernel}
    K(\vec{x},\vec{x'})=\rm{exp}(-\gamma||\vec x-\vec x'||^2),
\end{equation}
where $\gamma$ is a hyper-parameter which defines the width of the Gaussian distribution. The objective function maximized by the algorithm is the so-called \emph{soft margin}, which is characterized by $\nu$, another hyper-parameter associated with the penalty related to wrong labelling~\cite{scholkopf2000support}.

The model creation phase (see, for example, \cite{oneto2018model} for an introduction on the topic) is repeated every six hours: each time the collected historical data are randomly split in two subsets.
The first subset (\emph{training set}) contains $75\%$ of the data and it is used to train the model over the grid $(\gamma,\nu) \in  \{ 10^{-3}, 10^{-2}, \cdots, 10^2 \} \times \{ 10^{-3}, 10^{-2}, \cdots, 10^2 \}$. 
The latter subset (\emph{validation set}) contains the remaining $25\%$ of the data and it is used to select the best pair $(\gamma,\nu)$,%
\footnote{Although in principle a different value for the pair $(\gamma,\nu)$ is allowed in every model creation phase, the validation procedure has selected $(\gamma,\nu)=(0.1,0.1)$ for the whole duration of our experiment.}
which is used to recreate the model on the whole data set (\emph{training} $+$ \emph{validation}). 

%
%
%%%%%
\renewcommand{\arraystretch}{1.1}
\begin{table*}[t]
\centering
\caption{Malware Description}
\label{tab:apt}
\begin{adjustbox}{max width=1\textwidth}
\begin{tabular}{|c|c|c|c|c|c|c|c|}

\hline
\textbf{Name}        & \textbf{Category} & \textbf{APT 
}                         & \textbf{Codec} & \textbf{DNS rtype} & \textbf{Domain}            & \textbf{
Detection} & \textbf{$A$} \\ 
\hline

Pisloader

& Trojan RAT                                  &

Wekby

& Base32                                   & TXT                                                                                                                                                                  & local.it-desktop.com                      & 96\%                                                                                                                                                                      & 100\%                                                                                                                                                     \\ \hline

ISMDoor
      
& Trojan RAT                                  & GreenBug                                                                                                                                                       & Base64                                   & AAAA                                                                                                                                                                 & basnevs.com                               & 91\%                                                                                                                                                                      & 100\%                                                                                                                                                     \\ \hline

Denis 

& Trojan Backdoor                             &

OceanLotus

& Base64                                   & NULL                                                                                                                                                                 & z.teriava.com                             & 100\%                                                                                                                                                                     & 100\%                                                                                                                                                     \\ \hline

Carbanak

& Trojan Backdoor                             & 
Fin7
  & Custom                                        & TXT                                                                                                                                                                  & en.google4-ssl.com                        & 100\%                                                                                                                                                                     & 100\%                                                                                                                           \\ \hline
Cobalt Strike 

& Commercial Tool                             & CopyKittens                                                                                                                                                     & Custom                                        & TXT                                                                                                                                                                  & update.cisc0.net                          & 100\%                                                                                                                                                                     & 100\%                                                                                                   \\ \hline
Bondupdater 

& Trojan Powershell                           & 
OILRig  
& Custom                                   & A, TXT                                                                                                                                                               & withyourface.com                          & 100\%                                                                                                                                                                     & 100\%                                                                                                                                                                                                                                 \\ \hline
UDPoS 

& PoS Malware                                 & -                                                                                                                                                              & RC4                                      & A                                                                                                                                                                    &              
\urll{ns.service-logmeln.network} & 100\%                                                                                                                                                                 & 100\%                                                                                                                                                     \\ \hline
DNSpionage
& Trojan RAT                                  & -                                                                                                                                                              & Base32                                   & A                                                                                                                                                                    & microsoftonedrive.org                     & 100\%                                                                                                                                                                     & 100\%                                                                                                                                                     \\ \hline

\end{tabular}
\end{adjustbox}
\end{table*}
\renewcommand{\arraystretch}{1}
%%%%%
%
%

\subsection{Online Phase}\label{sec:online}

The online phase analyzes real time data and involves the following steps:

\begin{enumerate}
\setlength\itemsep{0em}
\item \emph{Feature extraction}: the same features described in Sect.~\ref{sec:offline} are extracted.
\item \emph{Classification}: the classifier trained in the offline phase is applied to online data in order to check whether they are consistent with the normal behavior; if they do not conform, the patterns are assessed with the subsequent module.
\item \emph{Advanced analytics}: an algorithm combines data mining techniques and cyber security knowledge to perform an in-depth analysis of the queries considered suspicious in the previous step. In the remainder of the section, we provide a description of this module.
\end{enumerate}

\subsubsection{Advanced analytics} The following anomaly indicators of possible covert channel activity are developed.

\emph{Number of unique requests from an IP address to a domain.} Avoiding the stub resolver cache during an attack may result in not-repeated requests to a domain. Therefore,
a large number of unique requests from an IP address to a domain may indicate a DNS covert channel. An indicator $i_r$ is defined as the frequency with which the number of unique requests from an IP address to a domain falls on the tails of its distribution (90th percentile).

\emph{Number of unique hostnames per domain.} Related to the previous point, also a large number of unique hostnames per domain might indicate a DNS covert channel.
An indicator $i_h$ is defined as the frequency with which the number of subdomains per domain falls on the tails of its distribution (90th percentile).

\emph{Entropy.} The entropy of a subdomain considered as a sequence of characters is related to its randomness.
Since attackers
compress the data to be sent as a payload via encoding or encryption, a large entropy may be a sign of an encoded or encrypted payload.
An indicator $i_e$ is defined as the maximum among the entropy of the whole subdomain and that of its longest label.

\emph{Distribution of frequencies of mono-grams and bi-grams.} A distribution of subdomain characters, which is distant from the distribution of characters of real languages, may represent another indicator of randomness~\cite{qi2013}, related to an encoded or encrypted payload.
For each considered language (English and Italian), $i_d^{\rm lang,mono}$ ($i_d^{\rm lang,bi}$) is defined as the maximum among the Jaro-Winkler distance~\cite{winkler1990string} of the mono-grams (bi-grams) distribution of the subdomain and that of its longest label from the corresponding distribution of the language. Finally, we evaluate the average $i_d= (i_d^{\rm eng,mono}+i_d^{\rm eng,bi}+i_d^{\rm ita,mono}+i_d^{\rm ita,bi})/4$.

\emph{Anomaly Index.}
Once the previously described indicators tailored to the DNS protocol are extracted, an \emph{anomaly index} $A$ is built by averaging them: $A=(i_r+i_h+i_e+i_d)/4$. 
For each machine source, the index is rescaled by taking into account the fraction between the number $n_s$ of suspicious queries and the total number $n_{\rm tot}$ of DNS queries produced, according to $A\to \rm{min}(1,A+b+c\,\rm{log}(n_s/n_{\rm tot}))$, where $(b,c)=(0.33,0.067)$ have been set with a preliminary \emph{una tantum} validation procedure.
The detection of covert channels has thus been reduced to a very simple one-dimension classification problem: only queries with $A>A_{\rm th}$ are labeled as suspicious, where the optimal threshold  ($A_{\rm th}=0.25$) has been found with the \emph{una tantum} validation procedure. Note that $A$ may be considered as an indicator of anomaly from a behavioral point of view.

%
%
%%%%%
\renewcommand{\arraystretch}{1.1}
\begin{table}[!ht]
%\large
\centering
\caption{Network Description}
\label{tab:network}
\begin{adjustbox}{max width=0.48\textwidth}
\begin{tabular}{|l|c|c|}
\hline
& \textbf{\makecell{15-Days \\Total}} & %%\textbf{One-Hour Average}
\textbf{\makecell{1-Hour \\Average}} 
\\ \hline

\textbf{N. of Machines}                    & 360                    & -                         \\ \hline
\textbf{N. of Client Machines}             & 346                    & -                         \\ \hline
\textbf{N. of Connections}                 & 43\,M                  & 287\,k                    \\ \hline
\textbf{N. of Resolved DNS Queries}        & 4\,M                   & 25\,k                     \\ \hline
\textbf{N. of Unique Resolved DNS Queries} & 119\,k                  & 791                      \\ \hline

\end{tabular}
\end{adjustbox}
\end{table}
\renewcommand{\arraystretch}{1}
%%%%%
%
%

%-------------------------------------------------------------------------------
\section{Experimental Evaluation}\label{sec:res}
%-------------------------------------------------------------------------------

The proposed DNS covert channel detection algorithm was evaluated over a real company network: in particular, the test set comprises 15 days of ordinary traffic %as described in Table~\ref{tab:network} 
-- described in Table~\ref{tab:network} -- 
with the injection of traffic which covers the most relevant exfiltration and tunneling
attacks. Note that the test set has been only used to test the performance of the algorithm and not to modify the algorithm or the parameters.

Two different experimental designs have been selected for the test of the algorithm, related with both \emph{exfiltration} and \emph{tunneling} attacks: a malware scenario and a tool scenario, that correspond to two different modes of injection described in Sect.~\ref{sec:malwareexfiltr} and \ref{sec:DNStunnel} respectively. Both scenarios have been recreated by injecting the malicious traffic into the ordinary traffic of a real network, %which is described 
described in Table~\ref{tab:network}.

\subsection{Malware Scenario}\label{sec:malwareexfiltr}

%removed packetot, copykittens2
The traffic related to malware attacks was injected by employing 8 pcaps -- collected from the public
sandboxes~\cite{hybrid,joesandbox,reverse} -- associated with 7 different APTs~\cite{copykittens1,ismdoor,pisloader,dnspionage,denisbackdoor,bondupdater,carbanak} and 1 PoS malware campaign~\cite{udpos}. Table~\ref{tab:apt} provides a brief description of each malware with the following information:

%
%
%%%%%
\begin{itemize}
  \setlength\itemsep{0em}
 	\item The malware name;
 	\item The category of the malware;
 	\item The name of the APT group related to the malware;
	\item The codec employed to encode the payload into the DNS requests;
	\item The list of DNS record types used by the malware;
	\item The domain present in each pcap, which is a known IoC associated to the malware;
	\item The percentage of detected occurrences, i.e., the number of detected queries divided by the total number of
	queries produced by the malware;
	\item The value of the anomaly index $A$, defined in Sect.~\ref{sec:cc}.
\end{itemize}
%%%%%
%
%

%
%
%%%%%
\renewcommand{\arraystretch}{1.1}
\begin{table*}[!t]
\centering
\caption{Tools Description}
\label{tab:tools}
\begin{adjustbox}{max width=1\textwidth}
\begin{tabular}{|c|c|c|c|c|c|c|}
\hline
\textbf{Name}            & \textbf{Platform}                                                                    & \textbf{Linux PenTest
Distro}               & \textbf{Codec}               & \textbf{DNS rtype} & \textbf{
Detection
} 
& \textbf{$A$} \\ \hline

\multirow{3}{*}{dnscat2} & \multirow{3}{*}{
Linux, Windows
}             

& \multirow{3}{*}{-}                                                
& \multirow{3}{*}{Hexadecimal} 

& TXT  & 100\%     & 100\%     \\ \cline{5-7}  &       &      &     & MX                                                        
& 100\%              & 100\%                                                            
\\ 
\cline{5-7} 
                         &                                                                                       &                                                                                       &                              & CNAME                                                              & 100\%                                                                   & 100\%                                                            \\ \hline
\multirow{2}{*}{dns2tcp} & \multirow{2}{*}{
Linux, Windows
}         
& \multirow{2}{*}{
Kali, BlackArch
}            

& \multirow{2}{*}{Base64}      & TXT                                                                & 100\%                                                                   & 100\%                                                            \\ \cline{5-7} 
                         &                                                                                       &                                                                                       &                              & KEY                                                                & 100\%                                                                   & 100\%                                                            \\ \hline
\multirow{6}{*}{iodine}  
& \multirow{6}{*}{\makecell{Linux, Windows, \\Mac OS X}
} & 
\multirow{6}{*}{
Kali, BlackArch, BackBox
} & \multirow{6}{*}{Base128}     & NULL                                                               & 70.08\%                                                                 & 100\%                                                            \\ \cline{5-7} 
                         &                                                                                       &                                                                                       &                              & TXT                                                                & 87.02\%                                                                 & 100\%                                                            \\ \cline{5-7} 
                         &                                                                                       &                                                                                       &                              & SRV                                                                & 85.03\%                                                                 & 100\%                                                            \\ \cline{5-7} 
                         &                                                                                       &                                                                                       &                              & MX                                                                 & 88.11\%                                                                 & 100\%                                                            \\ \cline{5-7} 
                         &                                                                                       &                                                                                       &                              & CNAME                                                              & 77.73\%                                                                 & 100\%                                                            \\ \cline{5-7} 
                         &                                                                                       &                                                                                       &                              & A                                                                  & 86.67\%                                                                 & 100\%                                                            \\ \hline
\multirow{3}{*}{DNScapy} & \multirow{3}{*}{
Linux, Mac OS X
}            & \multirow{3}{*}{-}                                                                    & \multirow{3}{*}{Base64}      & CNAME                                                              & 13.73\%                                                                 & 69\%                                                             \\ \cline{5-7} 
                         &                                                                                       &                                                                                       &                              & TXT                                                                & 12.61\%                                                                 & 71\%                                                             \\ \cline{5-7} 
                         &                                                                                       &                                                                                       &                              & TXT, CNAME                                                         & 14.75\%                                                                 & 76\%                                                             \\ \hline

dnsfilexfer              & 
\makecell{Linux, Windows, \\Mac OS X}
&
{
Kali, BlackArch
}
& Hexadecimal                  & A                                                                  & 100\%                                                                 & 100\%                             

\\ \hline
Your-Freedom              &                   
\makecell{Linux, Windows, \\Mac OS X}
&  -                                                                             & Base64                       & NULL                                                               & 
100\%                                                                  & 100\%                                                            \\ \hline

\end{tabular}
\end{adjustbox}
\end{table*}
\renewcommand{\arraystretch}{1}
%%%%%
%
%

%%%%%
\subsection{Tool Scenario}\label{sec:DNStunnel}

The traffic related to attacks performed by tools was injected in the
network described in Table \ref{tab:network}
by using 5 of the most
popular DNS-tunneling tools~\cite{det,dns2tcp,dnscapy,dnscat2,iodine,yourfreedom} and 1 DNS-file-transfer tool~\cite{dnsfilexfer}. This choice is motivated by the state-of-the-art in the exploitation of DNS protocol for covert channel, assessed by the presence of these tools inside some Linux distributions for penetration testing, the supported DNS record types and the state of maintenance of the tools. Table~\ref{tab:tools} provides a brief description of each tool with the following information:

%
%
%%%%%
\begin{itemize}
    \setlength\itemsep{0em}
 	\item The name of the tool;
 	\item The list of compatible platforms;
 	\item The Linux penetration testing distribution where the tool is pre-installed;
 	\item The codec employed to encode the payload into the DNS requests;
 	\item The list of DNS record types supported by the tool;
	\item The percentage of detected occurrences, i.e., the number of detected queries divided by the total number of queries produced by the malware;
	\item The value of the anomaly index $A$, defined in Sect.~\ref{sec:cc}.
\end{itemize}
%%%%%
%
%

All the tools considered in this paper require two different machines: (\emph{i}) a host which represents a server hold by an attacker and (\emph{ii})  another host which represents either an unaware infected client or a client controlled by an insider threat.
Both components have been recreated in an appropriate way in each experiment (see Table~\ref{tab:tools}).
\begin{enumerate}
    \setlength\itemsep{0em}
    \item The server component was simulated, whenever possible, with an appropriate version of a Linux penetration testing distribution such as Kali Linux~\cite{kali}, BlackArch Linux~\cite{blackarch}, and BackBox~\cite{backbox}.  In the remaining cases an \emph{ad hoc} Linux installation was employed;
    \item The client component was simulated with a Windows or Linux client, with an appropriate version of the operative system (for reasons of compatibility with installed tools).
\end{enumerate}

All tools employed in the experimental evaluation are open-source except Your-Freedom. A brief description of the tools considered follows.

\emph{iodine}~\cite{iodine} is one of the most popular DNS-tunneling tools~\cite{born2010,bubnov2018,homen2017detection,liu2017detecting,nadler2019detection} and is pre-installed in all Linux penetration testing  distributions considered in this paper. Iodine encapsulates an IPv4 packet into the payloads of DNS packets and needs a TUN/TAP device to operate. It uses the NULL record type by default, but can support other record types such as PRIVATE, TXT, SRV, MX, CNAME and A (returning CNAME). Upstream data is GZIP compressed and encoded; the supported encoding option includes Base32, Base64, Base64URL and Base128. If NULL or PRIVATE record types are used, downstream data is transmitted as GZIP compressed raw IP packet bytes; if other record types are used, it is GZIP compressed and encoded like upstream data.

\emph{dnscat2}~\cite{dnscat2,liu2017detecting,qi2013} is designed to create an encrypted command-and-control (C\&C) channel over the DNS protocol. It uses the TXT, CNAME and MX record types by default, but it supports also A and AAAA record types if data are only sent from client to server. All data in both directions is transported using hexadecimal encoding.

\emph{dns2tcp}~\cite{born2010,bubnov2018,dns2tcp,liu2017detecting} relays TCP connection over DNS, and is pre-installed in Kali Linux and in BlackArch Linux. Data encapsulation already takes place at the TCP level, so no separate driver (TUN/TAP) is required. Dns2tcp requires the list of available resources (e.g., ssh, smpt, pop3) on server configuration, i.e., the resources that the client can request to access. It uses the TXT record type by default, but it can also support the KEY record type. All data in both directions is transported using Base64 encoding.

\emph{DNScapy}~\cite{bubnov2018,dnscapy} creates a SSH tunnel through DNS packets. The idea of encapsulating SSH in DNS comes from OzymanDNS~\cite{ozyman}. DNScapy supports CNAME and TXT record types but the default mode is RAND, which randomly employs both CNAME and TXT. All data in both directions is transported using Base64 encoding.

\emph{dnsfilexfer}~\cite{dnsfilexfer} exfiltrates files via DNS lookup and is pre-installed in Kali Linux and in BlackArch Linux. Dnsfilexfer supports only the A record type. All data is transported using hexadecimal encoding.

\emph{Your-Freedom}~\cite{usenixinternet2013,ucaminternet2017,yourfreedom} is a 
tool based on a service available either in a paid version or in a free version, with some limitations. Only the client component can be downloaded and it requires either OpenVPN or a software that acts as a \emph{socksifier}. An appropriate online server, accessible by the client component, has to be chosen during the setup of the client. Your-Freedom supports many DNS record types such as NULL, WKS, TXT, CNAME and MX. All data in both directions is transported using Base64 encoding.

\subsection{Experimental Results}
The DNS covert channel detection algorithm described in Sect.~\ref{sec:cc} has been evaluated over a 15-day-long experimental session performed in
a test network (described in Table \ref{tab:network}) with the injection of traffic related to several malware and tool
attacks. Tables \ref{tab:apt} and \ref{tab:tools} (described in Sects.~\ref{sec:malwareexfiltr} and \ref{sec:DNStunnel} respectively) clearly show that the proposed method successfully detected all the covert channel attacks with high anomaly and detected occurrence indicators.
In particular, all tools described in Table \ref{tab:tools} have been %fully
detected with high detection rate with the exception of DNScapy. In this context, the main difference between DNScapy and the other variants is the fact that the former produces many state queries to maintain the established connection alive; these queries are shorter and are not identified by the algorithm, which focuses on exfiltration queries. Anyway, it is important to note that the proposed method is indeed able to detect a covert channel attack via DNScapy, as the exfiltration queries are correctly detected. The main purpose of the algorithm is therefore fully reached despite the fact that the detection rate on the queries is low for this particular tool.

In Table~\ref{tab:results} we summarize the performance of the algorithm via the confusion matrix, which contains:

\begin{itemize}
\setlength\itemsep{0em}

\item True Negatives %rate 
($T_N$): The number of unique legitimate queries that are correctly labeled as legitimate.

\item False Negatives %rate 
($F_N$): the number of unique queries related with malicious covert channels queries that are incorrectly labeled as legitimate.

\item False Positives %rate 
($F_P$): the number of unique legitimate queries that are incorrectly labeled as covert channels; many of them are related with particular types of advertising which produce queries that look very similar to covert channels\footnote{Here are some examples of false positive: (\emph{i}) "\emph{0w57c49k-db0dd2cc45455ae425c83e3b8ed8a67a14261606-am1.d.aa.online-metrix.net}", (\emph{ii}) "\emph{5b584d886b0f49f795209d5763d8c078.events.ubembed.com}", (\emph{iii}) "\emph{y2tyfol9hiuw5hzwe2hnusbzm1qz51545315830.nuid.imrworldwide.com}".}.

\item True Positives %rate 
($T_P$): the number of unique queries related with malicious covert channels that are correctly detected.

\end{itemize}

%
%
%%%%%
\renewcommand{\arraystretch}{1.1}
\begin{table}[!ht]
%\large
\centering
\caption{Results (Confusion Matrix)}
\label{tab:results}
\begin{adjustbox}{max width=0.48\textwidth}
% \usepackage{multirow}
\begin{tabular}{|c|c|c|c|c}
\cline{3-4}
\multicolumn{2}{l|}{\multirow{2}{*}{}} 
& \multicolumn{2}{c|}{\textbf{\makecell{Actual Class}}} &  \\ \cline{3-4}
\multicolumn{2}{c|}{}                    &\textbf{\makecell{Normal}}   & \textbf{\makecell{Malicious}}   &  \\ \cline{1-4}
\multirow{2}{*}{\begin{tabular}[c]{@{}c@{}} \textbf{\makecell{Predicted}} \\ \textbf{\makecell{Class}}\end{tabular}} & \textbf{\makecell{Normal}}    & $T_N=116649$    & $F_N=490$     &  \\ \cline{2-4}    & \textbf{\makecell{Malicious}} & $F_P=2033$      & $T_P=18174$   &  \\ \cline{1-4}
\end{tabular}
\end{adjustbox}
\end{table}
\renewcommand{\arraystretch}{1}
%%%%%
%
%

A remarkable result is the low rate of false negatives: this determines indeed a 97\% recall, also known as detection rate, $R=T_P/(T_P+F_N)$. In particular, false negatives are mainly due to the evasion of the state queries related to the DNScapy tool -- as previously explained in this Section -- while false positives are mostly related with advertising queries. 

Another metric commonly used to evaluate binary classifiers is the F-score. It is defined as $F=2\,P\,R/(P+R)$ (where $P=T_P/(T_P+F_P)$), and it is a more reliable indicator for problems in which the classes are highly unbalanced, as in the present case (the number of legitimate queries in the test set is much larger than the number of covert channel related queries). The value obtained in the experiments is $F=94\%$.
Even though an exact comparison is not possible due to the use of different datasets, we can note that we obtained an F-score almost as large as the one in the work of \cite{liu2017detecting}, despite some crucial differences: (\emph{i}) the method proposed in \cite{liu2017detecting} employs a supervised classifier, i.e., a binary classifier is trained using both legitimate and malicious queries, while our approach trains the classifier only with the normal network traffic; this means that our approach can identity a new variant of a DNS covert channel without a specific training for that variant;
(\emph{ii}) the benign parts of the training and test sets in \cite{liu2017detecting} only contain domains in the Alexa  top 1-million list \cite{alexa}, while our corresponding sets contain all the traffic collected from a real network.
We can therefore conclude that the proposed method brings a relevant contribution %improvement
in the state-of-the-art of DNS covert channel detection.

%-------------------------------------------------------------------------------
\section{Conclusions}\label{sec:concl}
%-------------------------------------------------------------------------------

In this paper, we proposed a DNS covert channel detection method based on the analysis of the DNS traffic of a single network; the analysis requires \emph{aramis} security monitoring system. The proposed framework employs a machine learning module which provides a behavioral analysis and specific anomaly indicators able to encompass the characteristics of a covert channel. The main contribution of this approach is the ability to provide network profiles tailored to the network users, and not to the single query events, hence allowing spotting possible deviations from the normal baseline. Moreover, models are created in an unsupervised mode, thus enabling the identification of zero-days attacks and avoiding the requirement of signatures or heuristics for new variants.

The proposed solution has been evaluated over a test network, with the injection of 8 pcaps associated with 7 different APTs and 1 PoS malware campaign and the network traffic of 5 DNS-tunneling tools and 1 DNS-file-transfer tool; all the malicious variants were detected, while producing a low false-positive rate during the same period. Another important contribution of the proposed method is therefore the capability to detect covert channels generated with a wider variety of malware and tools, compared with the state-of-the-art.

As a future development, we plan to extend the analysis to other protocols, (e.g., HTTP and HTTPS), and to refine the linguistic anomaly indicators by adding more languages or custom dictionaries. 
Moreover, we intend to broaden the behavioral approach in order to create another level of profiles tailored to groups of network users sharing common behavioral characteristics (e.g., same department, same job).

%-------------------------------------------------------------------------------

\bibliographystyle{latex8}
\bibliography{bibtex}

\end{document}